\documentclass[aps,twocolumn,prd,superscriptaddress,nofootinbib]{revtex4}
\usepackage[english]{babel}
\usepackage{epsfig}
\usepackage{psfrag}
\usepackage{graphicx}
\usepackage{amsmath}
\usepackage{amssymb}
\usepackage{dsfont}
\newcommand{\dfq}{\!\!\!\frac{d^4q}{(2 \pi)^4}}

\newcommand{\pslash}{p\hspace{-1.7mm}/}
\newcommand{\phslash}{\hat{p}\cdot\vec{\gamma}}
\newcommand{\pvslash}{\vec{p}\cdot\vec{\gamma}}

\newcommand{\wpslash}{{\omega}_{p}\gamma_{4}}
\newcommand{\beq}{\begin{equation}}
\newcommand{\eeq}{\end{equation}}

\newcommand{\eq}[1]{Eq.~(\ref{#1})}
\newcommand{\eqs}[1]{Eqs.~(\ref{#1})}

\begin{document}

\title{Color-spin locking in a self-consistent Dyson-Schwinger approach}


\author{Florian Marhauser}
\affiliation{Institut f\"ur Kernphysik, 
Technische Universit\"at Darmstadt,
D-64289 Darmstadt, Germany}

\author{Dominik Nickel}
\affiliation{Institut f\"ur Kernphysik, 
Technische Universit\"at Darmstadt,
D-64289 Darmstadt, Germany}

\author{Michael Buballa}
\affiliation{Institut f\"ur Kernphysik, 
Technische Universit\"at Darmstadt,
D-64289 Darmstadt, Germany}

\author{Jochen Wambach}
\affiliation{Institut f\"ur Kernphysik, 
Technische Universit\"at Darmstadt,
D-64289 Darmstadt, Germany}
\affiliation{Gesellschaft f{\"u}r Schwerionenforschung mbH, Planckstra{\ss}e
  1, D-64291 Darmstadt, Germany}


\date{\today}


\begin{abstract}
We investigate the color-spin locked (CSL) phase of
spin-one color-superconducting quark matter using a truncated Dyson-Schwinger 
equation for the quark propagator in Landau gauge. 
Starting from the most general parity conserving ansatz allowed by the CSL 
symmetry, the Dyson-Schwinger equation is solved self-consistently 
and dispersion relations are discussed. We find that chiral symmetry is 
spontaneously broken due to terms which have previously been 
neglected. As a consequence, the excitation spectrum contains only gapped 
modes even for massless quarks.
Moreover, at moderate chemical potentials the quasiparticle pairing 
gaps are several times larger than expected from extrapolated weak-coupling 
results. 
\end{abstract}


\maketitle


\section{Introduction}

In recent years, the possible existence of various color superconducting
phases in Quantum Chromo Dynamics (QCD) at low temperatures and sufficiently 
high densities has received much attention 
(for reviews see \cite{reviews,Rischke:2003mt}). 
At asymptotic densities, where the problem can be treated within a
weak-coupling expansion, there is compelling evidence \cite{weakCFL}
that quark matter is in the color-flavor locked (CFL) phase where up, down, 
and strange quarks participate in a scalar (spin 0) diquark condensate 
\cite{cfl}. 
The situation is much less clear at ``moderate'' densities of about 3 to
10 times nuclear matter density, which could be reached in the centers
of compact stars.  
These densities could be too small to allow for a sizeable fraction of
strange quarks and, hence, pairing of strange quarks with nonstrange
quarks might be disfavored.
On the other hand, the cross-flavor pairing of up and down quarks
in a two-flavor superconducting (2SC) phase
could be unfavorable as well if local neutrality and beta equilibrium
are taken into account \cite{absence2sc}.
Indeed, as demonstrated in various model calculations, BCS pairing of 
quarks with unequal flavors could be absent in the relevant regions
of the phase diagram \cite{Ruster:2005jc,Abuki:2005ms}.

In such environments quarks of different flavors might pair independently. 
In this case, if we restrict ourselves to pairing channels which are 
antisymmetric in color, spin-0 pairing is Pauli-forbidden and the most
promising candidates are Cooper pairs with total spin 1. 
Such phases  were discussed first in Refs.  
\cite{Schafer:2000tw,Alford:2002rz} and their properties were  
investigated in great detail in Ref.~\cite{Schmitt:2004et} for
weak-coupling. It was found that, among the various pairing patterns, 
the most stable spin-1 phase is the so-called color-spin locked 
(CSL) phase, where the three spin projections are locked to 
the three color degrees of freedom of the antitriplet.  
This is quite analogous to the B phase of superfluid 
helium-3, where the spin is locked to the orbital angular momentum. 

In Ref.~\cite{Aguilera:2005tg} the CSL phase has been analyzed 
within an Nambu-Jona-Lasinio (NJL)-type model. In this model, the locking of color and
spin was introduced through a gap matrix 
\beq
    \Phi^+ \;=\; \Delta\, \vec\gamma\cdot\vec I~,
\label{cslnjl}
\eeq
where $\vec\gamma = (\gamma_1,\gamma_2,\gamma_3)$ are the gamma matrices 
with spacelike indices and 
$\vec{I} = (\lambda_7, -\lambda_5, \lambda_2)$ 
are the antisymmetric Gell-Mann matrices, acting in color space. 
It was found that for massive quarks all quasiparticle modes are
gapped with the smallest gap in the spectrum being proportional to 
the constituent quark mass. 
This is an interesting result because ungapped quark matter in general 
leads to rapid neutron star cooling via the direct URCA process 
which is difficult to reconcile with the empirical data.
In fact, in a recent analysis of neutron star cooling data, 
it was concluded that in a possible quark core all quasiparticle modes 
should be gapped with the smallest gap of the order of 10-100 keV
and decreasing with density \cite{Grigorian:2004jq}.
The NJL-model results of Ref.~\cite{Aguilera:2005tg} turned out to
be in rough agreement with these constraints. 
However, the size of the smallest gap was found to be very
sensitive to the model parameters (see also Ref.~\cite{Aguilera:2006cj},
where a similar model with a nonlocal interaction was used).
Moreover, in order to obtain gaps which are decreasing with density
the authors used a density dependent coupling which had to be
introduced by hand.  
In this situation a more fundamental, QCD-based approach is clearly
desirable.

A second problem was raised in Ref.~\cite{Schmitt:2005wg}. 
Here the color-spin locking was introduced in a similar way as in
\eq{cslnjl} but with $\vec\gamma$ replaced by a transverse gamma matrix,
$\vec\gamma_{\perp}(\vec p) = \vec\gamma - (\vec\gamma\cdot\hat p)\,\hat p$,
where $\hat p = \vec p/|\vec p|$ and $\vec p$ is the 3-momentum dependence 
of the gap matrix.
In this case it turns out that there is always an ungapped mode, 
even for massive quarks.
Of course, the question whether or not there are ungapped modes in the
spectrum should not depend on an arbitrary ansatz for the gap matrix. 
Instead, the gap structure should be the result of a selfconsistent
dynamical calculation searching for the most favored ground state.
Again, this problem cannot be solved within NJL-type models,
because the point interaction does not allow for momentum dependent
structures, like $\vec \gamma_{\perp}$. 

Recently, there has been considerable progress in studying QCD at
nonzero chemical potential within the framework of Dyson-Schwinger
equations (DSEs) in Landau gauge \cite{NWA1,DN,NWA2}. Dyson-Schwinger 
equations provide a set of coupled integral equations that are the 
quantum equations of motion of the Green functions of a quantum field 
theory. If they are solved completely, the whole dynamics of the 
theory under consideration is understood.
Of course, in practice, one has to perform truncations. 
By now there is a lot of experience in setting up a truncation 
scheme for Landau gauge in the vacuum, whose results are in good agreement with lattice data
(for reviews see \cite{reviews2}). In Refs.~\cite{NWA1,NWA2} this scheme has 
been extended to describe spin-0 color superconductors at nonzero chemical 
potential. This has been done in such a way that at very large
chemical potentials
the weak-coupling results of Ref.~\cite{Wang:2001aq} are correctly
reproduced. To that end, medium modifications of the gluon propagator
through particle-hole excitations have been implemented.

Thus being constrained from ``two ends'' (lattice calculations in vacuum 
and weak-coupling results at very large $\mu$), the framework has then
been applied to the ``interesting'' regime of moderate chemical potentials.
Here the authors found considerable deviations from both the naively 
extrapolated weak-coupling pairing gaps \cite{NWA1} and the NJL-model
behavior of the dynamical quark masses \cite{NWA2}. 

Motivated by these results, we have extended the formalism of 
Refs.~\cite{NWA1,NWA2} to describe spin-1 color superconductors in the
CSL phase. 
Main focus is thereby the question about the smallest excitation gap, 
raised by the (partially contradicting) results of 
Refs.~\cite{Aguilera:2005tg} and \cite{Schmitt:2005wg}.
To avoid similar problems related to a specific ansatz for the 
gap function, we set up a fully self-consistent scheme, 
starting from the most general parity conserving CSL symmetric ansatz 
for the normal and anomalous quark self-energies.
Since in Ref.~\cite{Aguilera:2005tg} (constituent) quark masses were 
found to be important, we also include chiral symmetry breaking terms.
As we will see, these terms are in fact crucial and, to our surprise,
even survive in the chiral limit.

This work is organized in the following way: In section \ref{sect:theo_setup} 
we give a brief presentation of the theoretical framework for our 
self-consistent solution of the Dyson-Schwinger equation. 
The numerical results are presented in section 
\ref{sect:results}. 
In section \ref{sect:Dispersion_relations}
we present a physical interpretation in terms of 
dispersion relations. 
We summarize what we accomplished and conclude in 
section \ref{sect:conclusion}.

\section{Theoretical Setup}\label{sect:theo_setup}

\subsection{The truncated quark DSE at finite chemical potential}

We follow the scheme and notations presented in~\cite{NWA1}
to get a truncated, but closed DSE for the quark propagator at a finite
chemical potential within the Nambu-Gor'kov formalism.
The renormalized quark DSE in Landau gauge with appropriate
quark-wave function and quark-gluon vertex renormalization constants, $Z_{2}$
and $Z_{1F}$, respectively, is then given by 
\begin{eqnarray}
  \mathcal{S}^{-1}(p)&=& 
  Z_{2}\mathcal{S}_{0}^{-1}(p)+Z_{1F}\Sigma(p).
  \label{qdse}
\end{eqnarray}
$\mathcal{S}_{0}$ denotes the bare quark propagator, $\mathcal{S}$ the full quark propagator
and $\Sigma$ the quark self-energy.
These objects are $2 \times 2$ matrices in Nambu-Gor'kov space,
consisting of normal and anomalous components:
\begin{eqnarray}
\mathcal{S}_{0}(p)&=&
\left(
  \begin{array}{cc}
    S_{0}^{+}(p) & 0\\
    0 & S_{0}^{-}(p)
  \end{array}
\right),
\nonumber\\
\mathcal{S}(p)&=&
\left(
  \begin{array}{cc}
    S^{+}(p)& T^{-}(p)\\
    T^{+}(p)& S^{-}(p)
  \end{array}
\right),
\nonumber\\
\Sigma(p)&=&
\left(
  \begin{array}{cc}
    \Sigma^{+}(p)&\Phi^{-}(p)\\
    \Phi^{+}(p)&\Sigma^{-}(p)
  \end{array}
\right).
\end{eqnarray}
Inserting this into \eq{qdse} and formally solving for the 
components of the full propagator, we arrive at
\begin{alignat}{2}
  T^{\pm} &= &
   -&Z_{1F}\left(Z_{2}{S^{\mp}_{0}}^{-1}+Z_{1F}\Sigma^{\mp}\right)^{-1}
  \Phi^{\pm}S^{\pm},
  \label{fullT}
  \\
  {S^{\pm}}^{-1} &= & &Z_{2}{S^{\pm}_{0}}^{-1}+Z_{1F}\Sigma^{\pm}
  \nonumber\\
  && -&Z_{1F}^{2}\Phi^{\mp}\left(Z_{2}{S^{\mp}_{0}}^{-1}+Z_{1F}\Sigma^{\mp}
  \right)^{-1}\Phi^{\pm}.
  \label{fullS}
\end{alignat}
Each of these components are, in principle, matrices in flavor, color,
and Dirac space. However, since we only consider pairing of quarks
with equal flavor, the flavor structure is trivial, and we can 
consider each flavor separately.  

The bare quark propagator is given by
\beq
S_{0}^{+}(p)= (-i\omega_{p}\gamma_{4}-i\vec{p}\cdot\vec{\gamma}+m_{0})^{-1}
\eeq
and
$S_{0}^{-}(p) = -CS_{0}^{+}(-p)^{T}C$.
Here we have introduced the notation $\omega_{p}=p_{4}+i\mu$.
Note that we have dropped possible flavor indices for the quark mass 
$m_0$ and the chemical potential $\mu$, 
which are in general flavor dependent. 
On the other hand, we exclude ``color chemical potentials'', i.e.,
constant gluonic background fields,
because it can be shown by the equations of motion
that these vanish in the CSL phase.

Defining the gluon momentum by $k = p - q$, 
the normal self-energy $\Sigma^{+}$ and the gap functions $\Phi^{+}$ in 
our truncation are obtained via 
\begin{alignat}{2}
  \Sigma^{+}(p) &= & &\frac{Z_{2}^{2}}{Z_{1F}} \pi \int \dfq\,
  \gamma_{\mu} \lambda_{a} S^{+}(q) \gamma_{\nu} \lambda_{a}\alpha_{s}(k^{2})
  \nonumber\\
  &&& \hspace{14mm}\times\left(
    \frac{P^{T}_{\mu\nu}}{k^{2}+G(k)}+
    \frac{P^{L}_{\mu\nu}}{k^{2}+F(k)} \right),
  \label{Sigmap}
  \\
  \Phi^{+}(p) &= & -&\frac{Z_{2}^{2}}{Z_{1F}} \pi \int \dfq\,
  \gamma_{\mu} \lambda_{a}^{T} T^{+}(q) \gamma_{\nu} \lambda_{a}
  \alpha_{s}(k^{2})
  \nonumber\\
  &&& \hspace{14mm}\times\left(
    \frac{P^{T}_{\mu\nu}}{k^{2}+G(k)}+
    \frac{P^{L}_{\mu\nu}}{k^{2}+F(k)}
  \right).
  \label{Phip}
\end{alignat}
Furthermore, we have $\Sigma^-(p) = -C{\Sigma^+(-p)}^{T}C$ and 
$\Phi^{-}(p) = \gamma_{4}\left.\Phi^{+}(p)\right.^{\dagger}  \gamma_{4}$,
which guarantees the action to be real \cite{Rischke:2003mt,Bailin:1983bm}.
Together with \eqs{fullT} and (\ref{fullS}),
these expressions form a set of integral equations which have to be solved 
self-consistently.

$P^{T}_{\mu\nu}$ and $P^{L}_{\mu\nu}$ are the usual projectors in the
subspace orthogonal to the gluon momentum $k_{\mu}$. $P^{T}_{\mu\nu}$ is 
transverse to the 3-momentum $\vec k$, 
whereas $P^{L}_{\mu\nu}$ is longitudinal.
The functions $G(k)$ and $F(k)$ encode Landau damping and Debye screening
of the gluons. In our analysis this is not included self-consistently,
but for simplicity $F$ and $G$ are computed using bare quark propagators. 
The result has similarities with the hard-dense-loop approximation at 
asymptotically large densities. (For details, see Ref.~\cite{NWA1}.)

The function $\alpha_{s}(k^{2})$, being referred to as the strong running 
coupling, is the only input in our truncation scheme.
Analogous to previous investigations of the 2SC and CFL phase~\cite{NWA1,NWA2},
we employ two different couplings. One of theses running couplings, named
$\alpha_{\mathrm{I}}(k^{2})$, has been determined in DSE studies of the Yang-Mills
sector~\cite{Fischer:2003rp}. It will serve as a lower bound to the coupling
strength, since it underestimates chiral symmetry breaking significantly in
our vertex approximation.
The other running coupling considered is chosen to reproduce lattice QCD
results for the quark and gluon propagators in the vacuum for Landau
gauge~\cite{Bhagwat:2003vw} and will be labeled $\alpha_{\mathrm{II}}(k^{2})$. 
In order to vary the renormalization point for this coupling,
we exploit the multiplicative renormalizability of the quark DSE.

As stated above, taking into account the medium polarization, Debye screening 
and Landau  damping are included. Both, screening and damping of the 
interaction, increase with the interaction strength. As a consequence the 
generated gap functions are relatively insensitive to variations of 
$\alpha_{s}(k^{2})$. Due to this fortunate instance, our approach
has quite a predictive power despite the uncertainties in the effective 
low-energy quark interaction.

The renormalization constants are determined in the chirally broken vacuum.
Inserting the above expressions for the normal and anomalous self-energy
into the DSE, $Z_{1F}$ cancels out.
We determine the quark wave function renormalization
constant $Z_{2}$ and the renormalization constant $Z_{m}$, relating the
unrenormalized quark mass $m_{0}(\Lambda^{2})$ at a cutoff $\Lambda$ to the
renormalized mass $m(\nu)$ via
\begin{eqnarray}
  m_{0}(\Lambda^{2}) &=& Z_{m}(\nu^{2},\Lambda^{2})m(\nu),
\end{eqnarray}
by requiring
\begin{eqnarray}
  \left. S^{+}_{q}(p)\right|_{p^{2}=\nu^{2}} &=& -i\pslash+m(\nu)
\end{eqnarray}
at a renormalization scale $\nu$.
The mass dependence of the quark wave function
renormalization constant $Z_{2}$ turns out to be 
negligible in the considered mass
ranges. Therefore $Z_{2}$ is determined in the
chiral limit.


\subsection{General CSL symmetric ansatz}

As pointed out before, the truncated Dyson-Schwinger equation presented 
above corresponds to a coupled set of integral equations.
In order to find a self-consistent solution, we parameterize the normal 
and anomalous self-energies in terms of scalar functions 
(so-called dressing functions) and determine normal and anomalous 
propagators via \eqs{fullT} and  (\ref{fullS}). 
These are then inserted into \eqs{Sigmap} and (\ref{Phip}) to
recalculate the self-energies, and the whole procedure is iterated
until convergence is achieved.

For this strategy 
we need to ensure that the basis for the parameterization
of the self-energies and propagators
forms a closed set, i.e., that no further structures
are generated during the iteration. 
To that end, we must allow for the most general parameterization 
which is consistent with all symmetries that are left unbroken in
the considered phase, i.e., in our case, the CSL phase.

In the Introduction we loosely characterized the CSL phase as 
spin-1 pairing of quarks with equal flavor, where the three
spin projections are locked to the three color degrees of freedom.
We should now be more precise:
In the CSL phase, the total (spin $+$ orbital) angular momentum of the pair 
is locked to an $SO_{c}(3)$ subgroup of the $SU_{c}(3)$ color symmetry.
This means,
the CSL ground state remains invariant under transformations 
of the quark fields, generated by the sum ~\cite{Schafer:2000tw}
\beq
  \vec{G} = \vec{J} + \vec{I},
\eeq
where $\vec J = \vec S + \vec L$, 
$\vec{S} = \frac{i}{2} \gamma_5 \gamma_4 \vec{\gamma}$
is the spin operator, $\vec L = \vec r \times \vec p$
is the orbital angular momentum,
and $\vec{I}=(\lambda_7,-\lambda_5,\lambda_2)$ is the generator of
the above-mentioned $SO_{c}(3)$ subgroup in color space.
One can easily verify that the generators $\vec G$ satisfy the commutation 
relations
\begin{eqnarray}
  \left[G_{i},G_{j}\right] &=& i\,\epsilon_{ijk}\,G_{k},
\end{eqnarray}
corresponding to a residual $SU_{J+c}(2)$ symmetry.

For a CSL invariant ground state, all Green functions are also 
invariant under CSL transformations. 
Therefore, we require  for the normal and anomalous propagators
\begin{eqnarray}
  S^{+}(\vec{p},p_{4}) &=&
  \phantom{-}US^{+}(\vec{p},p_{4})\gamma_{4}U^{\dagger}\gamma_{4},
\nonumber\\
  T^{+}(\vec{p},p_{4}) &=& \phantom{-}UT^{+}(\vec{p},p_{4})C^{-1}U^{T}C,
\end{eqnarray}
where $U\in SU_{J+c}(2)$.
In addition, we restrict ourselves to even parity pairing, 
i.e., we require
\begin{eqnarray}
  S^{+}(\vec{p},p_{4}) &=&
  \phantom{-}\gamma_{4}S^{+}(-\vec{p},p_{4})\gamma_{4},
\nonumber\\
  T^{+}(\vec{p},p_{4}) &=& -\gamma_{4}T^{+}(-\vec{p},p_{4})\gamma_{4}.
\end{eqnarray}
We will now present a self-consistent parameterization under these
conditions.
Since only quarks with equal flavor are pairing in the CSL phase, the
different flavors decouple completely in our truncation. 
(In fact, the number of flavors only appears in the medium polarization, 
i.e., as a prefactor in the screening functions $F(k)$ and $G(k)$.)
As a consequence, we only need to parameterize the color-Dirac structure.
Nevertheless, the number of terms which are consistent with our
requirements -- and therefore have to be considered -- is still
very large:
We find that normal and anomalous self-energies  
as well as normal and anomalous propagators must be parameterized by
20 independent functions each.

The result can be written in the following form,
\begin{alignat}{3}
  S^{+}(p)
  &= & Z_{2} \sum_{i=1}^5\Big(
    &-i\,\pvslash\, S^{+}_{A,i}(p)-i\,\wpslash\, S^{+}_{C,i}(p) &&
\nonumber \\
  &&&+S^{+}_{B,i}(p)\quad -i\,\gamma_{4}\,\pvslash\, S^{+}_{D,i}(p) 
  &&\Big)~P_{i},
\nonumber\\
  T^{+}(p)
  &= & Z_{2} \sum_{i=1}^5\Big(
    &i\,\gamma_{4}\,\phslash\, T^{+}_{A,i}(p)+\,\gamma_{4}\,T^{+}_{B,i}(p) &&
\nonumber \\
  &&&+T^{+}_{C,i}(p)\qquad +i\, \phslash\, T^{+}_{D,i}(p) &&\Big)~M_{i},
\nonumber\\
  \Sigma^{+}(p)
  &= & \frac{Z_{2}}{Z_{1F}} \sum_{i=1}^5\Big(
    &-i\,\pvslash\,\Sigma^{+}_{A,i}(p)-i\,\wpslash\,\Sigma^{+}_{C,i}(p) &&
\nonumber \\
  &&&+\Sigma^{+}_{B,i}(p)\quad -i\,\gamma_{4}\,\pvslash\,\Sigma^{+}_{D,i}(p) 
  &&\Big)~P_{i},
\nonumber\\
  \Phi^{+}(p)
  &= & \frac{Z_{2}}{Z_{1F}} \sum_{i=1}^5\Big(
   &i\,\gamma_{4}\,\phslash\,\phi^{+}_{A,i}(p)+\,\gamma_{4}\,\phi^{+}_{B,i}(p) 
   &&
\nonumber \\
  &&&+\phi^{+}_{C,i}(p)\qquad +i\, \phslash\,\phi^{+}_{D,i}(p)
  &&\Big)~M_{i},
\label{Eq:ansatz_gap_fct}
\end{alignat}
where $P_i$ and $M_i$ are matrices in color-Dirac space
defined by
\begin{eqnarray}
  (P_1)_{ij} &=&  \delta_{ij}
\nonumber\\
  (P_2)_{ij} &=& -\gamma_{5}\left(\hat p\cdot \vec I_{ij}\right)
\nonumber\\
  (P_3)_{ij} &=& \gamma_{4}\gamma_5\left(\vec{\gamma}\cdot \vec
    I_{ij}\right)
\nonumber\\
  (P_4)_{ij} &=& \hat p_i \hat p_j
\nonumber\\
  (P_5)_{ij} &=& \gamma_{4}\left(\gamma_i \hat p_j+\hat p_i\gamma_j\right)
\end{eqnarray}
and $M_i = \gamma_5 \gamma_4 P_i$. The indices $i,j$ are indices in color 
space while the Dirac structure is implicit.
As before, $\hat{p}=\vec{p}/\vert\vec{p}\vert$.

This ansatz can be derived by considering the largest linear independent
basis in color-Dirac space under the $SU_{J+c}(2)$ symmetry.
It can furthermore be deduced by coupling two quark fields in three internal
$SU(2)$ representations to an overall singlet.
Our notation of factorizing the right hand sides of \eq{Eq:ansatz_gap_fct}
into four pure Dirac terms and the color-Dirac matrices $P_i$ or $M_i$
is mainly motivated by the structure of the propagators and self-energies 
in a normal-conducting medium,
which are unit matrices in color space ($\equiv P_1$) and are customarily
parameterized by the same Dirac terms.  
Also note that the functions  $S^{+}_{F,i}(p)$, $T^{+}_{F,i}(p)$
$\Sigma^{+}_{F,i}(p)$, and $\phi^{+}_{F,i}(p)$
are renormalization point independent. 
Moreover, those terms with $F=A$ or $F=C$ are chirally symmetric,
whereas the terms with $F=B$ or $F=D$ break chiral symmetry. 
We want to emphasize, that the tensor structure is completely
determined by the residual symmetry and therefore independent of our
truncation.

The defining equation for the anomalous propagator, Eq.~(\ref{fullT}), is
homogeneous.
Therefore we are free to choose a global phase for the
gap function. In our calculations, we chose this phase, such that
\begin{eqnarray}
  \phi^{+}_{F,i}(p_{4},\vert\vec{p}\vert) &=&
  \phi^{+}_{F,i}(-p_{4},\vert\vec{p}\vert)^{*}
\end{eqnarray}
and therefore, all scalar gap functions are real for $p_{4}=0$.

Having discussed the most general ansatz for the CSL phase under even-parity
pairing, we can map out previous non-self-consistent ans\"atze by their
specific
parameterization of the gap functions as given in 
Eq.~(\ref{Eq:ansatz_gap_fct}).
For the ansatz used in Ref.~\cite{Aguilera:2005tg} the mapping is particularly
simple. It is given by
\begin{eqnarray}
  \phi_{C,3}^+ &=& \Delta
\label{njlansatz}
\end{eqnarray}
and all other scalar gap functions vanish. 

A more elaborate ansatz has been chosen in
Refs.~\cite{Schmitt:2004et,Rischke:2003mt,Schafer:2000tw}.
In addition, there was also a distinction between different cases of the CSL 
phase. One of these cases, the so-called ``transverse case'' maps into our 
notation by
\begin{eqnarray}
  \phi^+_{A,2} = \phi^+_{C,3} = \frac{1}{2} (\phi^+ + \phi^-),
\nonumber\\
  \phi^+_{A,3} = \phi^+_{C,2} = \frac{i}{2} (\phi^+ - \phi^-),
  \label{Eg:FFM_transverse_ansatz}
\end{eqnarray}
with $\phi^{\pm}$ being the (anti-)quasiparticle gap functions in the notation 
of Refs.~\cite{Schmitt:2004et,Rischke:2003mt}.
In weak-coupling calculations $\phi^{-}$ is set to zero, since it does not 
contribute to $\phi^{+}$ even to subsubleading order \cite{Rischke:2003mt}.

A second ansatz discussed in Refs.~\cite{Schmitt:2004et,Rischke:2003mt}
is called the ``mixed case'' and breaks chiral symmetry. 
For this case the mapping is
\begin{eqnarray}\label{Eg:FFM_mixed_ansatz}
  \phi^+_{A,2} = \phi^+_{C,3} = \phantom{-} \phi^+_{B,2} =\frac{1}{2} (\phi^+
  + \phi^-),
\nonumber\\
  \phi^+_{A,3} = \phi^+_{C,2} = -\phi^+_{D,2} = \frac{i}{2} (\phi^+ - \phi^-).
\end{eqnarray}
In the weak-coupling limit, where again $\phi^{-}$ is set to zero,
this ansatz is energetically disfavored compared to the ansatz given by 
Eq.~(\ref{Eg:FFM_transverse_ansatz}).

\section{Results}
\label{sect:results}

\subsection{The scalar gap functions as functions of the chemical potential}

We begin with the numerical results for
the gap functions $\phi^+_{F,i}$ at the Fermi surface,
i.e., at $p_{4}=0$ and $\vert\vec{p}\vert=p_{F}$. The Fermi
momentum $p_F$ corresponds to the momentum where the gap functions
have their maxima.\footnote{In general, the Fermi momenta can be defined via Luttinger's theorem,
see Ref.~\cite{NWA1}. In the present case, however, the much simpler
approximation as being located at the maxima of the gap functions works extremely well.
As the gaps are relatively small, the anomalous propagator is strongly
peaked at the Fermi surface and, consequently, also the gap function. The
error due to this approximation is small compared with the shift of the
Fermi momenta due to the normal self-energy.}   
We restrict our discussion to the gap functions 
$\phi^+_{F,2}$ and $\phi^+_{F,3}$, which correspond to the
attractive channels.
The other gap functions are induced by self-consistency, but their 
size turns out to be negligible.

\begin{figure}
  \includegraphics[width=8.0cm]{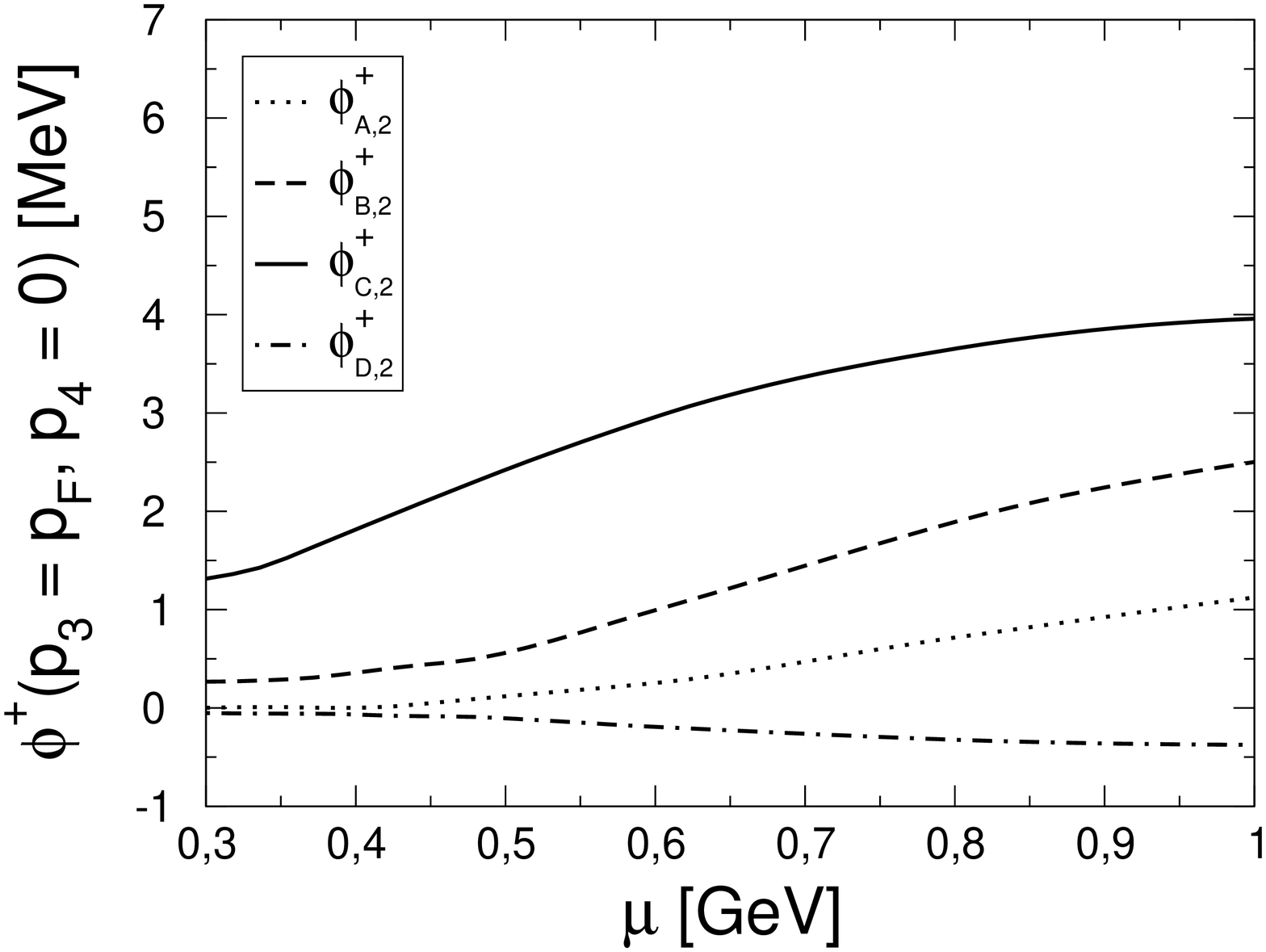}
  \includegraphics[width=8.0cm]{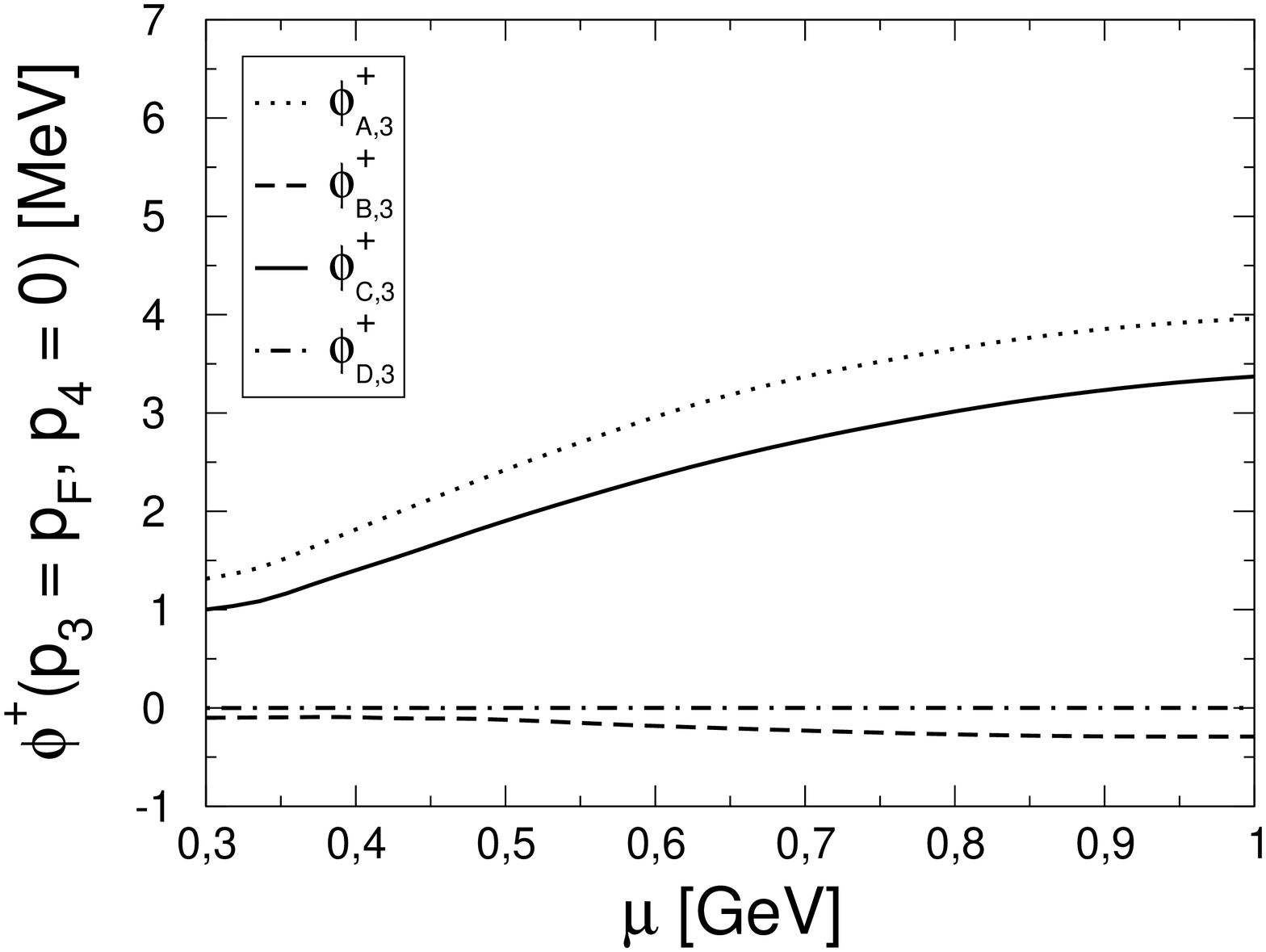}
  \caption{The values of the gap functions $\phi^+_{F,2}$ (upper panel)
    and $\phi^+_{F,3}$ (lower panel), $F=A,B,C,D$, at the
    Fermi surface for quarks with a renormalized mass of
    18.4 MeV at the renormalization point $\nu = 2$ GeV, obtained with 
    the strong running coupling $\alpha_{\mathrm{I}}(k^{2})$.}
  \label{Fig:massive_gaps_2and3_structure_dse}
\end{figure}

\begin{figure}
  \includegraphics[width=8.0cm]{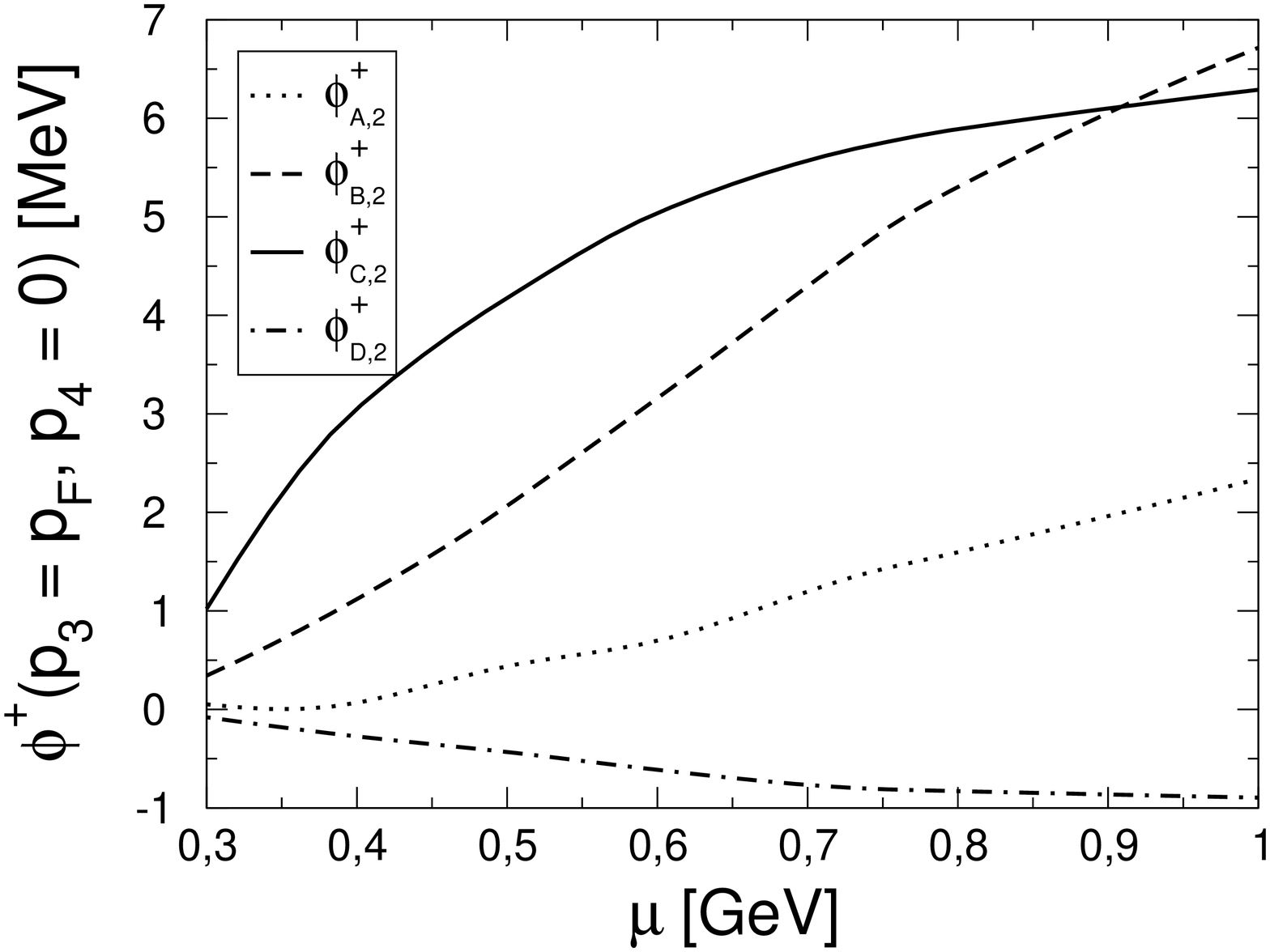}
  \includegraphics[width=8.0cm]{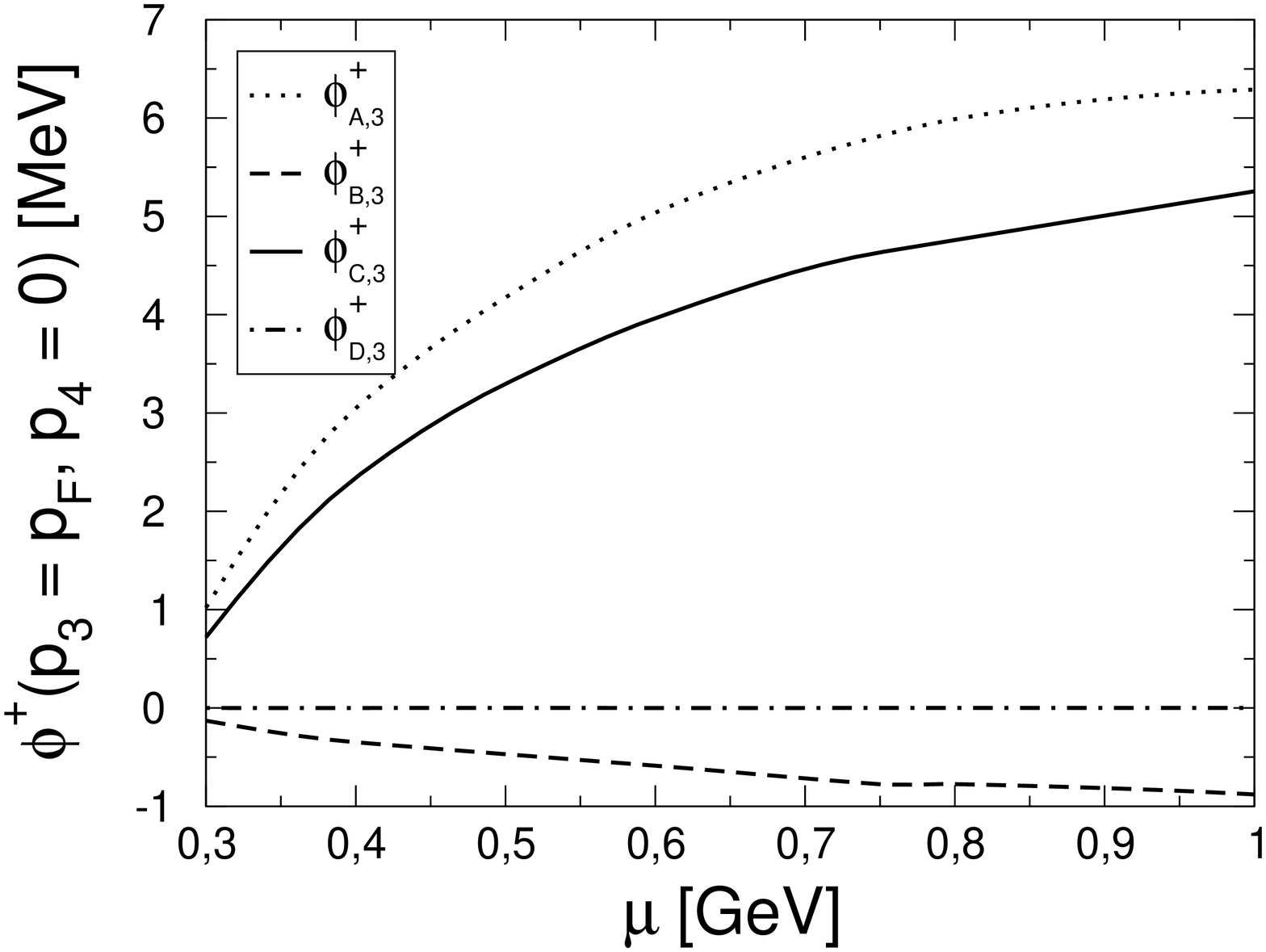}
  \caption{The same as Fig.~\ref{Fig:massive_gaps_2and3_structure_dse},
           but for the running coupling $\alpha_{\mathrm{II}}(k^{2})$ with a 
           renormalized mass of 18 MeV at the renormalization point $\nu = 2$ GeV.} 
  \label{Fig:massive_gaps_2and3_structure_latt}
\end{figure}

Our results as functions of the chemical potential $\mu$ are shown in
Fig.~\ref{Fig:massive_gaps_2and3_structure_dse},
which has been obtained
with the (weaker) strong running coupling $\alpha_{\mathrm{I}}(k^{2})$,
and in Fig.~\ref{Fig:massive_gaps_2and3_structure_latt}, which
corresponds to $\alpha_{\mathrm{II}}(k^{2})$.
As one might expect, there is a general tendency that the 
chiral symmetry conserving gap functions, 
$\phi^+_{A,i}$ and $\phi^+_{C,i}$, are more important than the
the symmetry breaking ones, $\phi^+_{B,i}$ and $\phi^+_{D,i}$.
One important exception is the function $\phi^+_{B,2}$ which can
be as big as or even bigger than the symmetry conserving functions. 
On the other hand, $\phi^+_{A,2}$ is relatively small. 

If we compare this with previous works, we see that the NJL ansatz, 
\eq{njlansatz}, is a rather poor approximation. 
But also the ``transverse case'',
\eq{Eg:FFM_transverse_ansatz}, misses the effect of $\phi^+_{B,2}$.
Only in the ``mixed case'', \eq{Eg:FFM_mixed_ansatz} all important
gap functions are taken into account, although not as independent
quantities. 
In fact, it is interesting to note that, within numerical precision,
we find that $\phi^+_{C,2} = \phi^+_{A,3}$, in agreement with
\eqs{Eg:FFM_transverse_ansatz} and 
(\ref{Eg:FFM_mixed_ansatz}).\footnote{This seems to indicate that this 
identity is protected by an underlying symmetry. However, until now, we 
have not been able to identify it.}
The other identities in these equations, however, are not found to hold. 

The scalar gap functions are of the order of a few MeV.
This is about two orders of magnitude larger than one would have expected
from extrapolating the weak-coupling results of the CSL phase to small 
chemical potentials. This remarkable result can be traced back to two 
features:
First, already for spin-0 phases, our approach leads to gap functions 
which are one order of magnitude larger than the extrapolated weak-coupling 
results~\cite{NWA1}.
And second, the suppression of the of the spin-1 gaps relative 
to the spin-0 gap is also about one order of magnitude less than 
in the weak-coupling limit.
E.g., for the ``transverse case'',
the latter is given by a factor of $e^{-9/2}$, which is the product of two
equal suppression factors of $e^{-9/4} \approx 0.1$ related to longitudinal 
and transverse gluons, respectively \cite{Schmitt:2004et}.
However, when we switch off the longitudinal gluons by hand, we find
that their contribution is almost negligible in our approach 
and, therefore, the suppression relative to the spin-0 result is only 
about one order of magnitude.

Comparing Fig.~\ref{Fig:massive_gaps_2and3_structure_dse}
with Fig.~\ref{Fig:massive_gaps_2and3_structure_latt}, 
we find that the different scalar gap functions respond  
differently to the change of the strong running coupling.
Typically, the results differ by a factor of 2, 
but the effect is somewhat smaller for $\phi^+_{C,2},\phi^+_{A,3}$ and 
$\phi^+_{C,3}$, whereas for $\phi^+_{B,2}$ it is about a factor of 2.5
at $\mu = 1$~GeV and even a factor of 4 at $\mu = 0.5$~GeV.
This is similar to the quark massfunction in the vacuum, which differs 
by a factor of 2-3 between the two couplings \cite{Fischer:2003rp}.


\subsection{The influence of the current quark mass}

\begin{figure}
  \includegraphics[width=8.0cm]{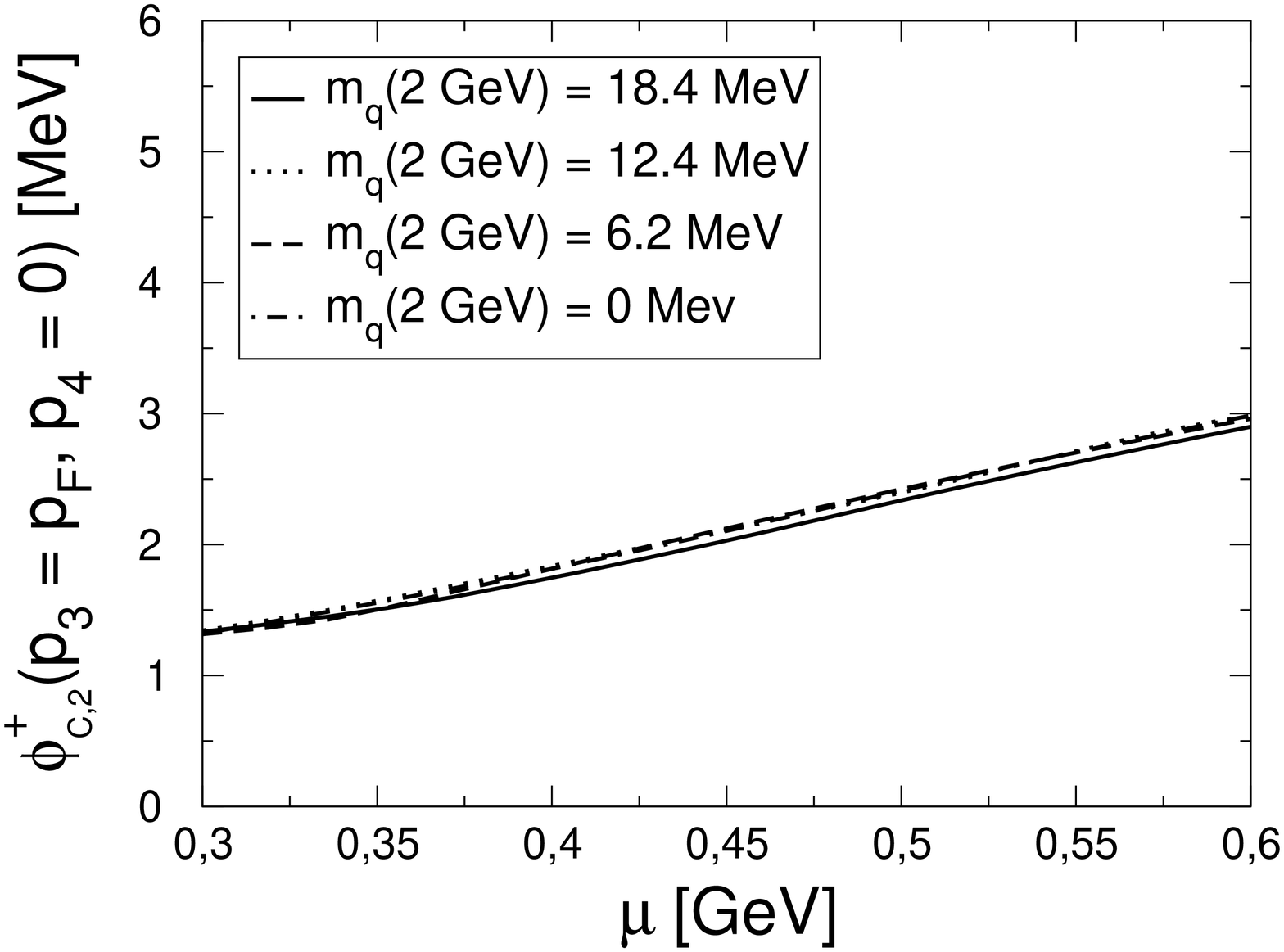}
  \includegraphics[width=8.0cm]{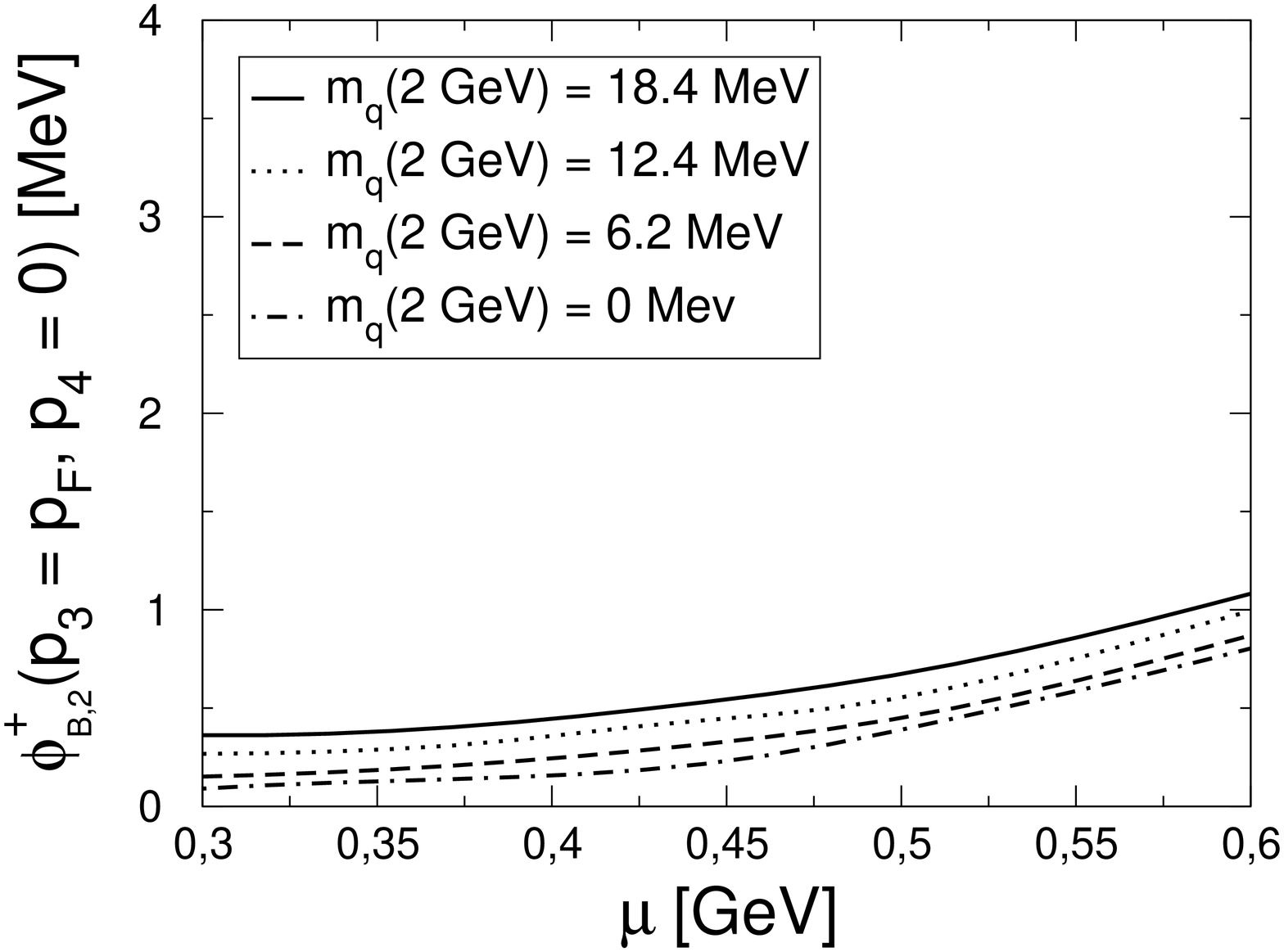}
  \caption{The values of the gap functions $\phi^+_{C;2}$
    (upper panel) and $\phi^+_{B;2}$ (lower panel)
    at the Fermi surface for various quark masses using the strong
    running coupling $\alpha_{\mathrm{I}}(k^{2})$.}
  \label{Fig:phic2b2ma1}
\end{figure}

\begin{figure}
  \includegraphics[width=8.0cm]{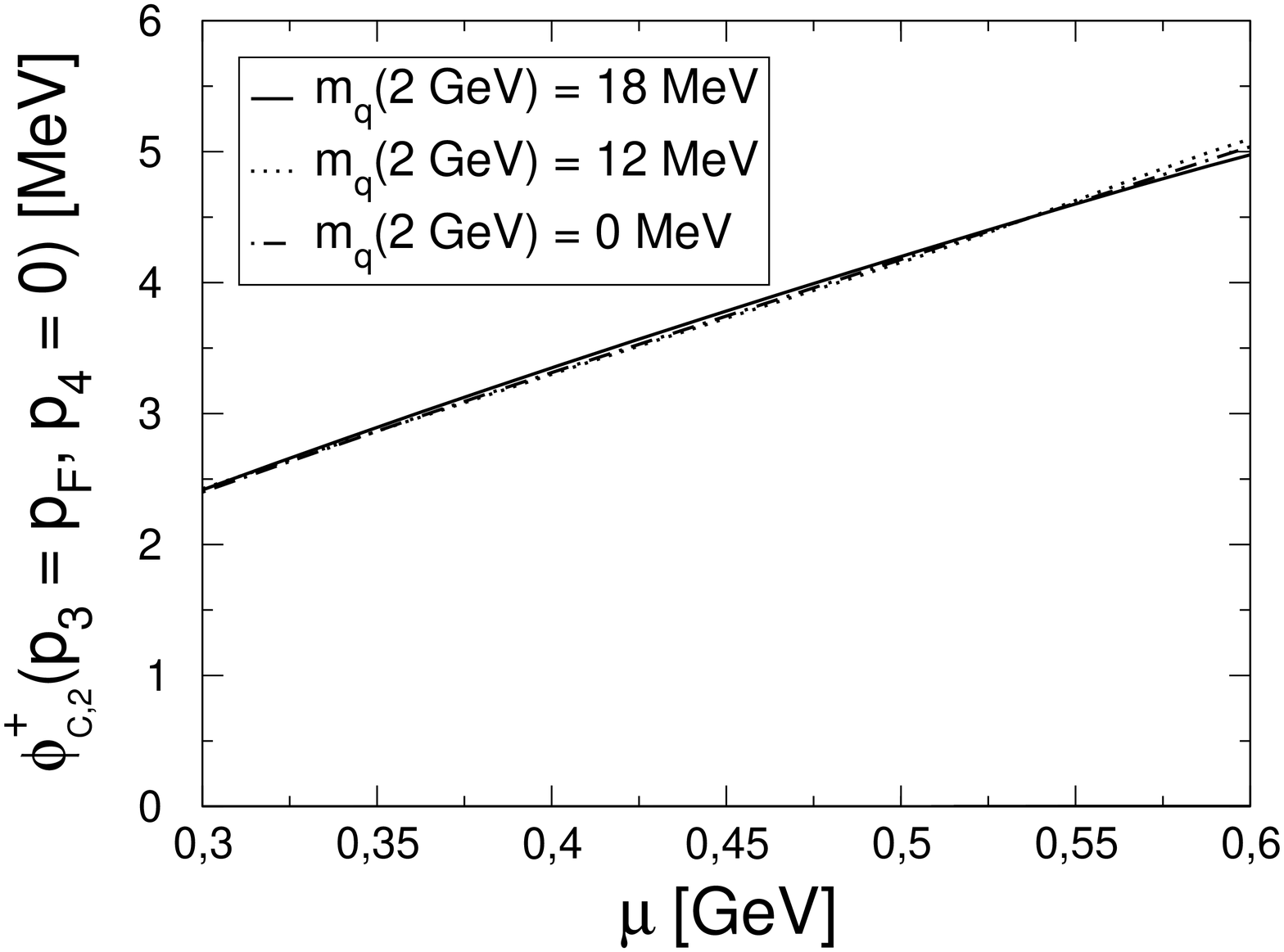}
  \includegraphics[width=8.0cm]{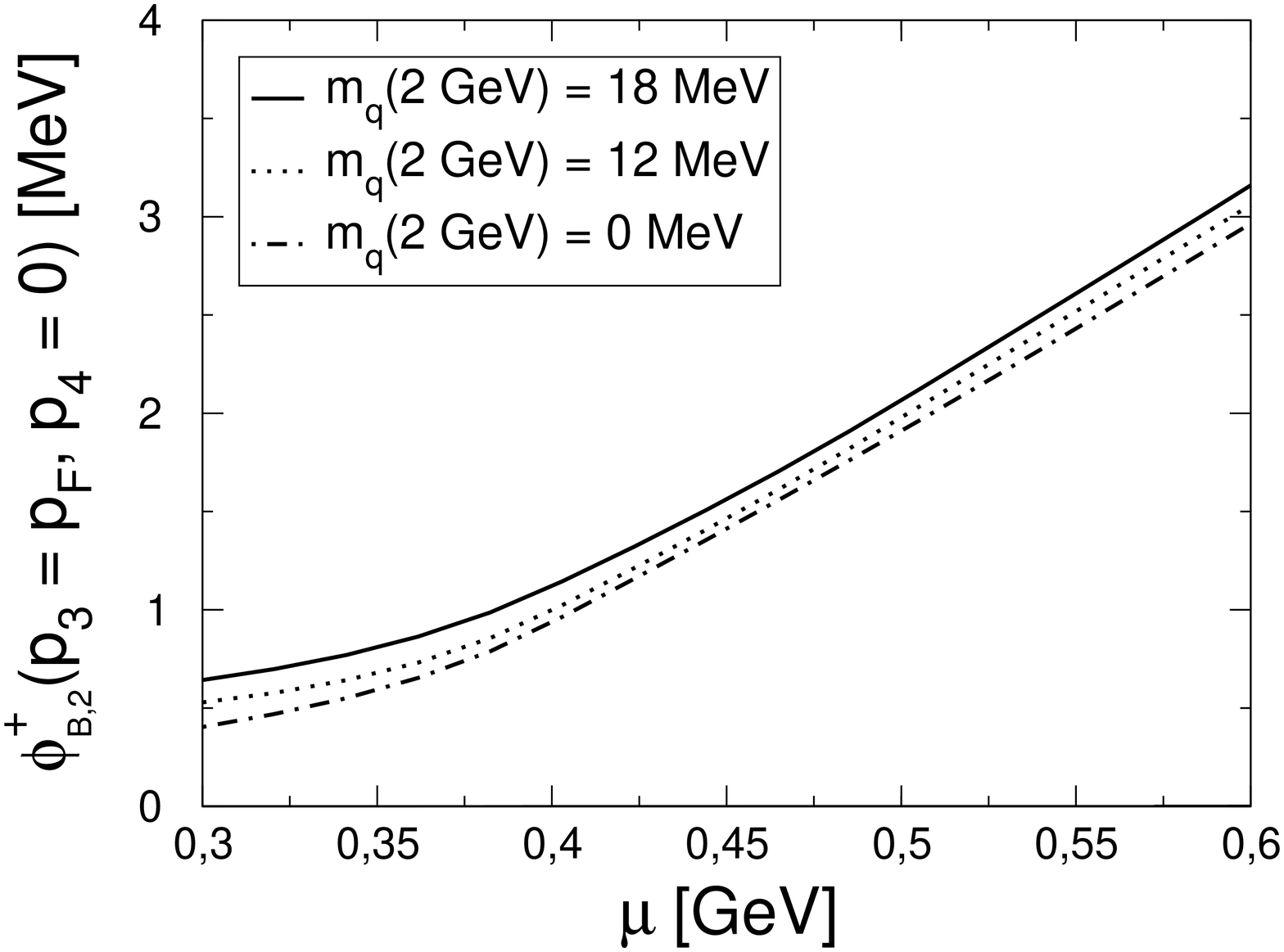}
  \caption{The same as Fig.~\ref{Fig:phic2b2ma1}, but using the running 
    coupling $\alpha_{\mathrm{II}}(k^{2})$ and slightly different masses.}
  \label{Fig:phic2b2ma2}
\end{figure}

Since quark masses played an important role in the NJL-model study
of Ref.~\cite{Aguilera:2005tg}, we have also investigated the effect 
of quark masses on the scalar gap functions. 
In Figs.~\ref{Fig:phic2b2ma1} and \ref{Fig:phic2b2ma2}
the values of $\phi^+_{C,2}$ and $\phi^+_{B,2}$ at the Fermi surface
are shown as functions of the quark chemical potential $\mu$
for various quark masses.
Obviously, the gap function $\phi^+_{C,2}$ is almost independent 
of the quark mass, whereas there is a moderate mass dependence of  
$\phi^+_{B,2}$.
Indeed, recalling that $\phi^+_{B,i}$ and $\phi^+_{D,i}$
break chiral symmetry, it was our expectation that these gap functions
are induced by nonvanishing quark masses and 
therefore the greater sensitivity of $\phi^+_{B,2}$ to the
quark mass appears rather natural. 
To our surprise, however, $\phi^+_{B,2}$ remains finite even for 
vanishing quark masses. 
This means, chiral symmetry is spontaneously broken by the
diquark condensates.
As we will see in the next section, this has important
consequences for the quasiparticle excitation spectrum. 

\begin{figure}
  \includegraphics[width=8.0cm]{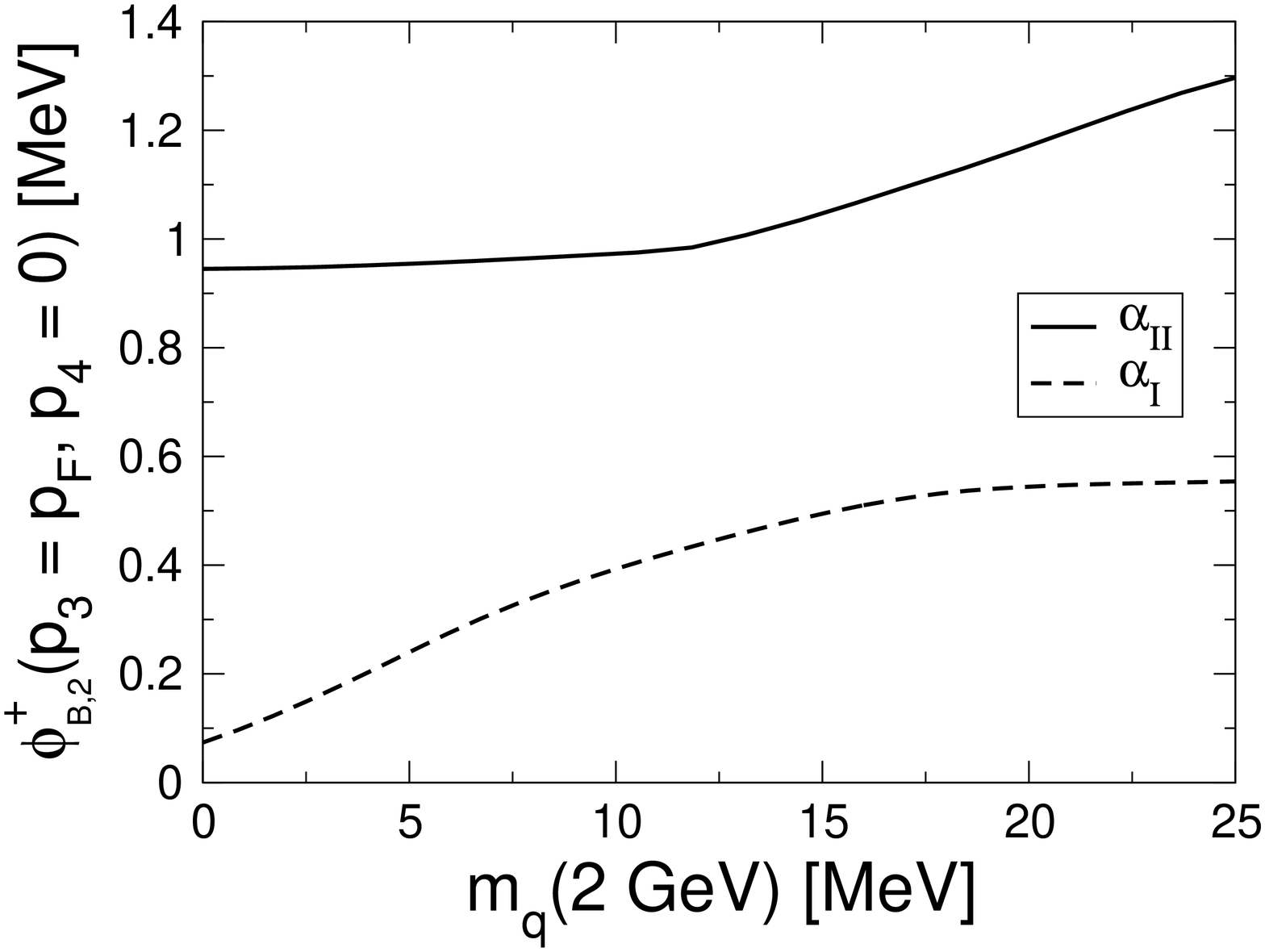}
  \includegraphics[width=8.0cm]{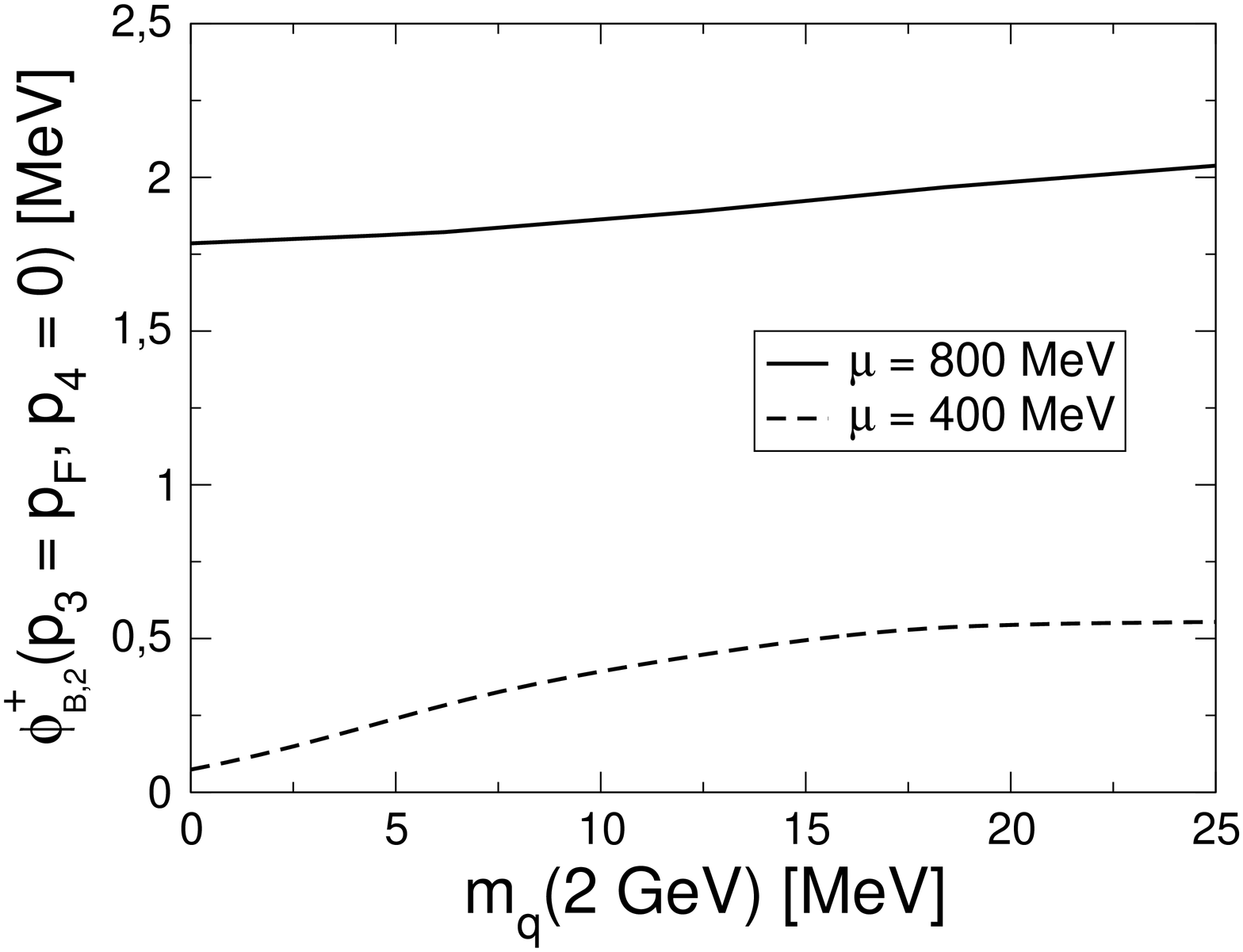}
  \caption{The values of the gap function $\phi^+_{B;2}$ 
    at the Fermi surface as a function of the renormalized quark mass
    at the renormalization point $\nu = 2$ GeV. Upper panel:
    Results for $\mu = 400$~MeV and different running couplings.
    Lower panel: Results for the running coupling $\alpha_{\mathrm{I}}(k^{2})$ 
    and different chemical potentials.}
  \label{Fig:gaps_for_var_m}
\end{figure}

The results of Figs.~\ref{Fig:phic2b2ma1} and \ref{Fig:phic2b2ma2}
suggest that the sensitivity of $\phi^+_{B,2}$ to the quark mass is more 
pronounced for the weaker coupling $\alpha_I(k^2)$ than for the stronger 
coupling $\alpha_{\mathrm{II}}(k^2)$. 
In order to analyze this behavior more systematically, 
the value of $\phi^+_{B,2}$ at the Fermi surface is shown 
in the upper panel of Fig.~\ref{Fig:gaps_for_var_m} as a function
of the quark mass for fixed $\mu = 400$~MeV and both running couplings.
For the stronger coupling $\alpha_{\mathrm{II}}(k^2)$, we find that 
$\phi^+_{B,2}$ is almost independent of the quark mass
for $m_q \lesssim 12$~MeV and then becomes more sensitive to the 
mass (dashed line). This is quite plausible:
At low quark masses, the value of $\phi^+_{B,2}$ is dominantly 
determined by the spontaneous symmetry breaking, whereas at
higher masses the explicit breaking plays a bigger role.

For the weaker coupling $\alpha_{\mathrm{I}}(k^2)$, on the other hand,
the spontaneous symmetry breaking part is much smaller and we find
a sizeable mass dependence right from the beginning (solid line). 
In fact, in this case the mass dependence becomes weaker at larger values
of $m_q$. 
This could be due to the fact that at fixed chemical potential
the density decreases with the quark mass. 
Note that $m_q$ corresponds to the quark mass at the
renormalization point $\nu = 2$~GeV and the effective masses at 
the scale of the chemical potential could be considerably larger.

The effect of increasing the chemical potential is very similar
to the effect of increasing the coupling strength. 
This can be seen in the lower panel of Fig.~\ref{Fig:gaps_for_var_m} 
where the value of $\phi^+_{B,2}$ at the Fermi surface is shown
as a function of the quark mass for  the coupling $\alpha_{\mathrm{I}}(k^2)$ 
and two different chemical potentials.
The dashed line corresponds to $\mu = 400$~MeV and is identical to
the dashed line in the upper panel. Increasing the chemical potential
to 800~MeV (solid line) we find again that the spontaneous symmetry 
breaking becomes much stronger and, hence, $\phi^+_{B,2}$ becomes
less sensitive to the quark mass.

\section{Dispersion relations}\label{sect:Dispersion_relations}

\subsection{Discussion}

Since the CSL phase might be relevant for compact star phenomenology, and in
particular the question whether there is an ungapped mode in the excitation
spectrum, it will be interesting to see how the dispersion relations behave
for a self-consistent solution. 
As mentioned in the Introduction, the dispersion relations for massive quarks 
in the CSL phase have  been investigated in \cite{Aguilera:2005tg} and
\cite{Schmitt:2005wg} with contradicting results:
The authors of Ref.~\cite{Schmitt:2005wg} used a DSE
approach in the weak-coupling limit and found that there is a gapless mode. 
In the NJL-model investigation of Ref.~\cite{Aguilera:2005tg}, on the
other hand, it was found that
there are only gapped modes in the excitation spectrum and that the smallest
effective gap is proportional to the constituent quark mass.
Since both investigations lack self-consistency, we have the chance to
gain more insight into the  dispersion relations with our self-consistent
ansatz. 

To describe the quasiparticle excitation spectrum, in principle
we have to extract the spectral functions. 
This is a relatively difficult task in our approach because one needs to 
analytically continue the results obtained in Euclidean space into 
Minkowski space. As demonstrated in Ref.~\cite{DN} for the case of
spin-0 color superconductors, this problem can be solved by applying
the maximum entropy method (MEM) \cite{MEM}.

In the following, however, we will pursue a simplified analysis,
aiming at a more qualitative discussion. 
To that end, we concentrate on the dispersion relations
$\varepsilon(\vert\vec{p}\vert)$ that are given as the solutions of the 
algebraic equation
\beq
  \mbox{Det} [\mathcal{S}^{-1}(\vec p,p_4)]\big|_{ip_4=\varepsilon(\vec p)} 
  = 0\,.
\label{Eq:disp_rel_def}
\eeq
Since our calculations are performed in Euclidean space, we can not access 
the roots of this equation. Therefore we apply the following approximations:
First, we neglect the normal self-energy of the quarks and second, we
replace the scalar gap functions in $\Phi^{\pm}(p)$ 
by their values at the Fermi surface.

Despite these simplifications, the resulting dispersion relations 
are extremely long expressions and difficult to handle.
Therefore we have searched for the solutions numerically. 
The results will be presented in the next subsection. 

However, to get a deeper understanding of the 
underlying mechanisms, it is instructive to perform further approximations 
which allow us to derive analytical expressions for the dispersion relations.
In particular, we wish to explore the significance of the gap function
$\phi^+_{B,2}$, which we found to be non-negligible in the self-consistent
solutions discussed in section \ref{sect:results}.
Led by these results, we consider a simple ansatz with 
\beq
 \phi^+_{A,3} = \phi^+_{C,2} \equiv \phi_{C}, \qquad
 \phi^+_{B,2} \equiv \phi_B,
\eeq
and neglect all other gap structures. The resulting dispersion relations
\footnote{Note, that in this simplified case $p_{F}=\mu$ as we neglect normal
self-energy contributions and quark masses.} are 
\begin{eqnarray}
  \varepsilon(|\vec p|) &=& \pm \sqrt{ (\mu \pm |\vec p|)^2 + \phi_B^2 }
\end{eqnarray}
and 
\beq
  \varepsilon(|\vec p|) =
  \pm \sqrt{ (\mu \pm |\vec p|)^2 + 
    \frac{1}{4} \left( \phi_B \pm \sqrt{\phi_B^2 + 8 \phi_C^2} \right)^2 }.
\eeq
Hence, the simultaneous presence of the gap $\phi^+_{B}$, representing 
opposite chirality pairing, and the gap $\phi^+_{C}$, representing equal 
chirality pairing, implies that all modes are gapped. 
This is similar to Ref.~\cite{Schmitt:2004et}, where the spectrum was 
found to be fully gapped in the ``mixed case'', \eq{Eg:FFM_mixed_ansatz}, 
but not in the ``transverse case'', \eq{Eg:FFM_transverse_ansatz}.


\begin{figure}
  \includegraphics[width=8.0cm]{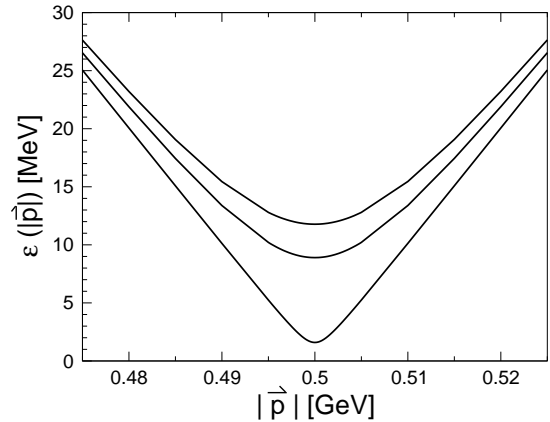}
  \caption{The quasiparticle dispersion relations for
  quarks with a renormalized mass of 18~MeV at the
  renormalization point $\nu = 2$~GeV obtained with
  the strong running coupling 
  $\alpha_{\mathrm{II}}(k^{2})$ at $\mu = 500$~MeV.}
  \label{Fig:disprel_for_var_mu}
\end{figure}

\subsection{Numerical results}

\begin{figure}
  \includegraphics[width=8.0cm]{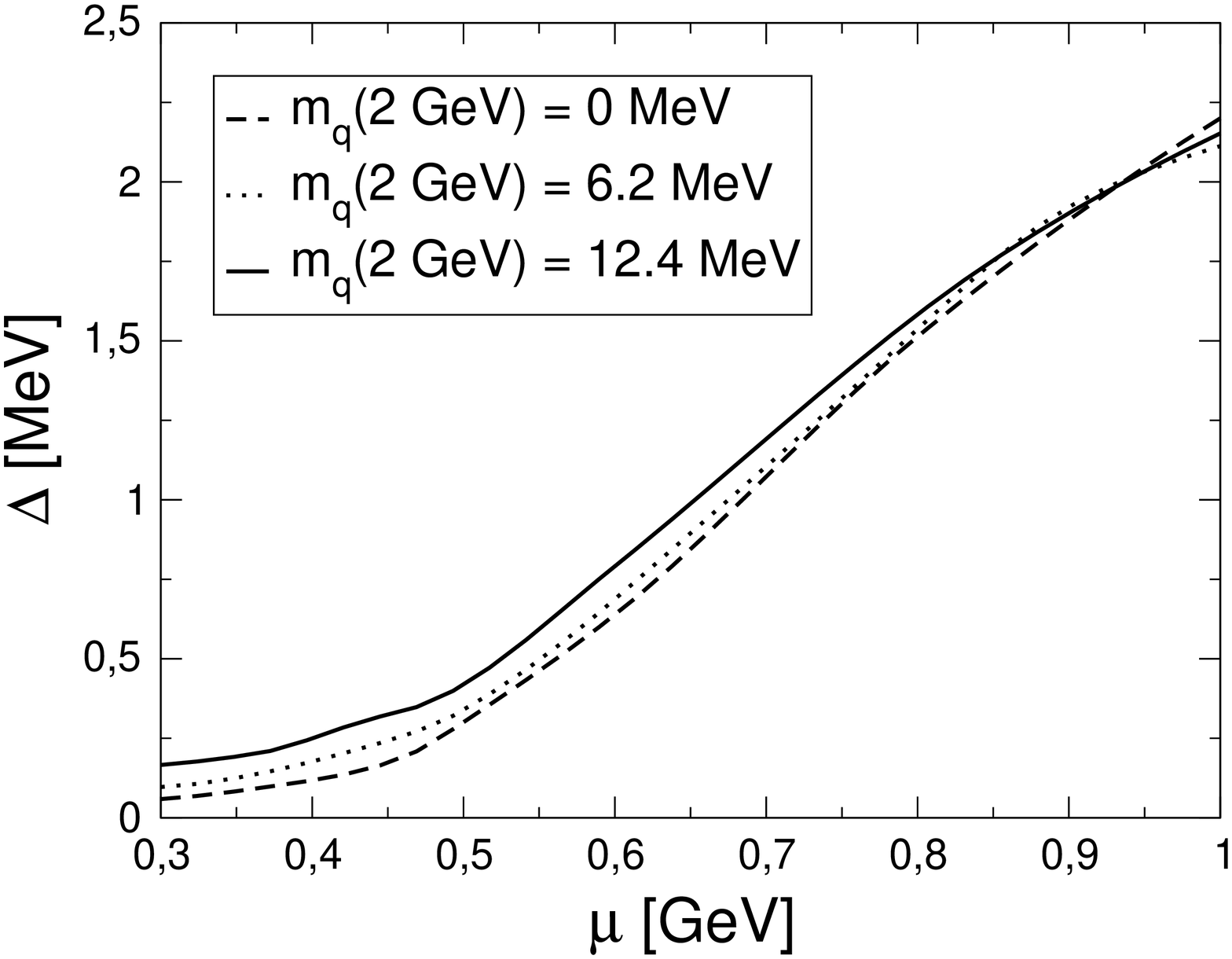}
  \includegraphics[width=8.0cm]{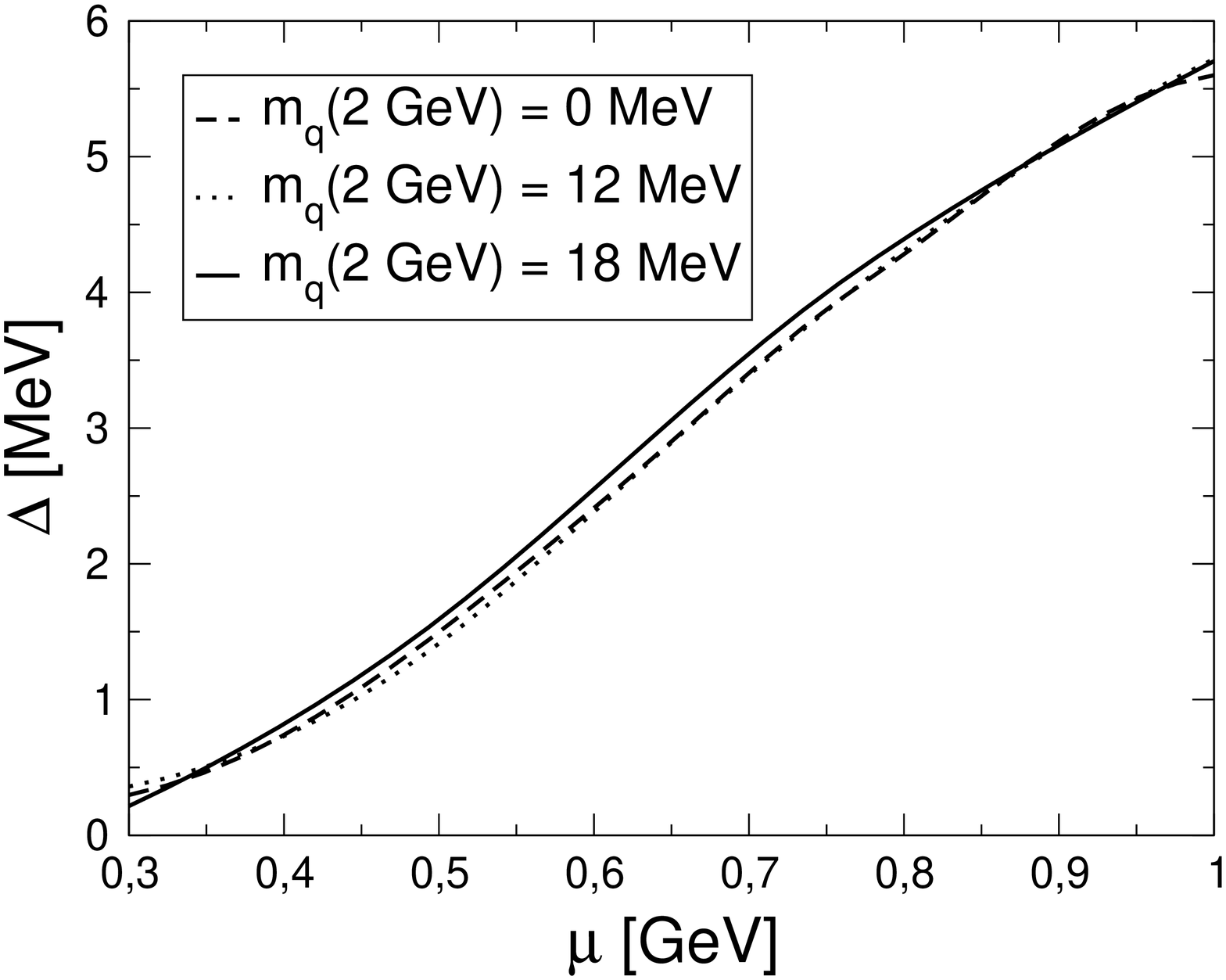}
  \caption{The smallest quasiparticle energy gaps 
    for massive quarks obtained with the
    running coupling $\alpha_{\mathrm{I}}(k^{2})$ (top) 
    and $\alpha_{\mathrm{II}}(k^{2})$ (bottom).}
  \label{Fig:disprel_smallest_gap}
\end{figure}

After these simplified considerations, we now return to the numerical 
solutions of \eq{Eq:disp_rel_def}, using the more complete, although
still approximate, expression for the inverse propagator.
Since $\phi^+_{B,2} \neq 0$ in the self-consistent solution of the 
Dyson-Schwinger equations, we expect that all quasiparticle modes
will be gapped.

The resulting dispersion relations at $\mu = 500$~MeV 
for massive quarks with a renormalized 
mass of $m_{q}(2~\mathrm{GeV})=18$~MeV are shown in
Fig.~\ref{Fig:disprel_for_var_mu}.
As expected, we observe only gapped modes in the excitation spectrum.
Of special interest for compact star cooling phenomenology is the 
behavior of the size of the smallest effective gap as a 
function of chemical potential.
This is shown in Fig.~\ref{Fig:disprel_smallest_gap} for the effective strong
running couplings $\alpha_{\mathrm{I}}(k^{2})$ and $\alpha_{\mathrm{II}}(k^{2})$ and for various
quark masses.
Since we find $\phi^+_{B,2}$ to be nonvanishing in the chiral limit,
corresponding to a spontaneous symmetry breaking, there are only gapped modes 
in the excitation spectrum, even for vanishing quark mass .
In addition we observe that the dependence of the effective gap on the quark
masses is weak, particularly for higher values of the chemical potential.
Although the value of the smallest gap is sensitive to the coupling, we can
conclude, that the gapfunction is monotonically increasing with the chemical
potentials and is of the order of a few 100~keV in the
astrophysically relevant regime.

\section{Summary and conclusions}\label{sect:conclusion}

We have studied the quark propagator in the color-superconducting CSL phase at
zero temperature in a self-consistent Dyson-Schwinger approach.
Our main objective has been the clarification of contradicting predictions by
previous investigations, especially concerning the phenomenologically
interesting size of the smallest excitation gap.
To this end, we have considered the most general ansatz for the quark
propagator allowed by the CSL symmetry pattern.
For comparison, we have outlined the truncated ans\"atze of previous
investigations and how they are parameterized in our basis.
In our numerical exploration, we found that the CSL phase exhibits a
color-spin structure that is close to the color-spin structure used in 
weak-coupling calculations~\cite{Schmitt:2004et}.
As a remarkable difference, chiral symmetry breaking terms are additionally
dynamically generated, leading to a spontaneous symmetry breakdown.
As a result, in contrast to~\cite{Schmitt:2005wg}, the size of the smallest
excitation gap is nonvanishing and triggered by a chiral symmetry breaking
diquark condensate.
Therefore, the dynamical mass generation, as elaborated
in~\cite{Aguilera:2005tg}, is not necessary for obtaining a nonvanishing
smallest excitation gap.
Although the value of the smallest gap is sensitive to the coupling, we can
conclude, that the gapfunction is increasing monotonically with the chemical
potentials and is of the order of a few 100~keV in the
astrophysically relevant regime.
This seems to be in conflict with the analysis of Ref.~\cite{Grigorian:2004jq}
where it was concluded that the smallest gap should decrease with the
chemical potential in order to describe the neutron star cooling 
data~\cite{Grigorian:2004jq}.
However, before drawing strong conclusions, a more careful analysis is
certainly required, where both specific heat and neutrino emissivities,
are derived within our approach.

\section{Acknowledgment}
We thank R. Alkofer, D. Rischke, A. Schmitt and  I. Shovkovy for helpful
discussions.

This work has been supported in part by the Helmholtz Association
(Virtual Theory Institute VH-VI-041) and by the BMBF under grant number
06DA916.




\begin{thebibliography}{99}

\bibitem{reviews} K.~Rajagopal and F.~Wilczek,
hep-ph/0011333;
M.~Alford,
Ann.\ Rev.\ Nucl.\ Part.\ Sci.\  {\bf 51}, 131 (2001);
T.~Sch{\"a}fer,
hep-ph/0304281;
M.~Buballa,
Phys.\ Rep.\ {\bf 407}, 205 (2005);
H.-C.~Ren,
hep-ph/0404074;
M.~Huang,
Int.\ J.\ Mod.\ Phys.\ E {\bf 14}, 675 (2005);
I.~A.~Shovkovy,
Found.\ Phys.\  {\bf 35}, 1309 (2005).

\bibitem{Rischke:2003mt}
D.~H.~Rischke,
Prog.\ Part.\ Nucl.\ Phys.\  {\bf 52}, 197 (2004).

\bibitem{weakCFL} I.~A.~Shovkovy and L.~C.~R.~Wijewardhana,
Phys.\ Lett.\ B {\bf 470}, 189 (1999);
T.~Sch\"{a}fer,
Nucl.\ Phys.\ {\bf B575}, 269 (2000).

\bibitem{cfl} M.~G.~Alford, K.~Rajagopal, and F.~Wilczek,
Nucl.\ Phys.\ {\bf B537}, 443 (1999).

\bibitem{absence2sc} 
M.~Alford and K.~Rajagopal,
JHEP {\bf 0206}, 031 (2002).

\bibitem{Ruster:2005jc}
S.~B.~R\"uster, V.~Werth, M.~Buballa, I.~A.~Shovkovy and D.~H.~Rischke,
Phys.\ Rev.\ D {\bf 72}, 034004 (2005).

\bibitem{Abuki:2005ms}
H.~Abuki and T.~Kunihiro,
Nucl.\ Phys.\ A {\bf 768}, 118 (2006).

\bibitem{Schafer:2000tw}  
T.~Sch\"afer,  
Phys.\ Rev.\ D {\bf 62} (2000) 094007.  
  
\bibitem{Alford:2002rz}  
M.~G.~Alford, J.~A.~Bowers, J.~M.~Cheyne and G.~A.~Cowan,  
Phys.\ Rev.\ D {\bf 67} (2003) 054018.  

\bibitem{Schmitt:2004et}
A.~Schmitt,
Phys.\ Rev.\ D {\bf 71} (2005) 054016.

\bibitem{Aguilera:2005tg}
D.~N.~Aguilera, D.~Blaschke, M.~Buballa and V.~L.~Yudichev,
Phys.\ Rev.\ D {\bf 72}, 034008 (2005).

\bibitem{Grigorian:2004jq}
H.~Grigorian, D.~Blaschke and D.~Voskresensky,
Phys.\ Rev.\ C {\bf 71}, 045801 (2005).

\bibitem{Aguilera:2006cj}
D.~N.~Aguilera, D.~Blaschke, H.~Grigorian and N.~N.~Scoccola,
Phys.\ Rev.\ D {\bf 74}, 114005 (2006).

\bibitem{Schmitt:2005wg}
A.~Schmitt, I.~A.~Shovkovy and Q.~Wang,
Phys.\ Rev.\ D {\bf 73}, 034012 (2006).

\bibitem{NWA1}
D.~Nickel, J.~Wambach and R.~Alkofer,
Phys.\ Rev.\ D {\bf 73}, 114028 (2006).

\bibitem{DN}
D.~Nickel,
Annals Phys. in print,
arXiv:hep-ph/0607224.

\bibitem{NWA2}
D.~Nickel, R.~Alkofer and J.~Wambach,
arXiv:hep-ph/0609198.

\bibitem{reviews2}
R.~Alkofer and L.~von Smekal,
Phys.\ Rept.\  {\bf 353}, 281 (2001);
C.~S.~Fischer,
J.\ Phys.\ G {\bf 32}, R253 (2006).

\bibitem{Wang:2001aq}
Q.~Wang and D.~H.~Rischke,
Phys.\ Rev.\ D {\bf 65}, 054005 (2002).

\bibitem{Bailin:1983bm}
D.~Bailin and A.~Love,
Phys.\ Rept.\  {\bf 107}, 325 (1984).

\bibitem{Fischer:2003rp}
C.~S.~Fischer and R.~Alkofer,
Phys.\ Rev.\ D {\bf 67}, 094020 (2003).

\bibitem{Bhagwat:2003vw}
M.~S.~Bhagwat, M.~A.~Pichowsky, C.~D.~Roberts and P.~C.~Tandy,
Phys.\ Rev.\ C {\bf 68}, 015203 (2003).

\bibitem{MEM}
M.~Jarrell and J.~E.~Gubernatis, 
Phys.\ Rept.\  {\bf 269}, 133 (1996);
M.~Asakawa, T.~Hatsuda and Y.~Nakahara,
Prog.\ Part.\ Nucl.\ Phys.\  {\bf 46}, 459 (2001).



\end{thebibliography}
\end{document}